\documentclass{article}

\usepackage{arxiv}

\usepackage[utf8]{inputenc} 
\usepackage{amsthm}
\usepackage{amssymb}
\usepackage{amsmath}
\usepackage[T1]{fontenc}    
\usepackage{hyperref}       
\usepackage{url}            
\usepackage{booktabs}       
\usepackage{amsfonts}       
\usepackage{nicefrac}       
\usepackage{microtype}      
\usepackage{lipsum}		
\usepackage{graphicx}
\usepackage{natbib}
\usepackage{xcolor}
\usepackage{algorithm}
\usepackage{algorithmic}
\usepackage{doi}
\usepackage{comment}
\hypersetup{breaklinks=true}
\title{SPaRSe-TIME: Saliency-Projected Low-Rank Temporal Modeling for Efficient and Interpretable Time Series Prediction}
\date{}

\renewcommand{\headeright}{}
\renewcommand{\undertitle}{}

\author{
\href{https://orcid.org/0009-0008-2571-7485}{\includegraphics[scale=0.06]{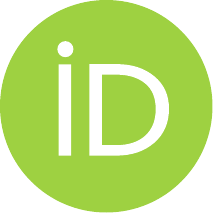}\hspace{1mm}K.~A.~Shahriar} \\
Department of Electrical and Electronic Engineering \\
Bangladesh University of Engineering and Technology \\
Dhaka-1205, Bangladesh \\
\texttt{kh.ashikshahriar@gmail.com}
}

\renewcommand{\shorttitle}{SPaRSe-TIME: Saliency and Low-Rank Temporal Modeling}

\hypersetup{
pdftitle={SPaRSe-TIME: Saliency-Projected Low-Rank Temporal Modeling for Efficient Time Series Prediction},
pdfauthor={K. A. Shahriar},
pdfkeywords={Time Series, Saliency, Low-Rank Modeling}
}

\begin{document}
\maketitle

\begin{abstract}
Time series forecasting is traditionally dominated by sequence-based architectures such as recurrent neural networks and attention mechanisms, which process all time steps uniformly and often incur substantial computational cost. However, real-world temporal signals typically exhibit heterogeneous structure, where informative patterns are sparsely distributed and interspersed with redundant observations. This work introduces \textbf{SPaRSe-TIME}, a structured and computationally efficient framework that models time series through a decomposition into three complementary components: saliency, memory, and trend. The proposed approach reformulates temporal modeling as a projection onto informative subspaces, where saliency acts as a data-dependent sparsification operator, memory captures dominant low-rank temporal patterns, and trend encodes low-frequency dynamics. These components are integrated through a lightweight, adaptive mapping that enables simplified, selective, and interpretable temporal reasoning. Extensive experiments on diverse real-world datasets demonstrate that SPaRSe-TIME achieves competitive predictive performance compared to recurrent and attention-based architectures, while significantly reducing computational complexity. The model is particularly effective in structured time series with clear temporal components and provides explicit interpretability through component-wise contributions. Furthermore, analysis reveals both the strengths and limitations of decomposition-based modeling, highlighting challenges in highly stochastic and complex multivariate settings. Overall, SPaRSe-TIME offers a principled alternative to monolithic sequence models, bridging efficiency, interpretability, and performance, and providing a scalable framework for time series learning.
\end{abstract}


\section{Introduction}

Time series data is ubiquitous, arising in domains ranging from financial markets \cite{praveen2026financial}, weather systems \cite{park2026towards}, resource optimization \cite{wen2022robust}, energy consumption \cite{DEB2017902}, healthcare \cite{crabtree1990individual}, and environmental monitoring \cite{zaini2022systematic}. By analyzing temporal dependencies and patterns, time series data enables forecasting of future values- commonly through regression (value prediction) \cite{han2019review}, classification (trend/direction prediction) \cite{Ismail_Fawaz2019-ui}, or anomaly detection (unexpected event identification) \cite{han2019review}. Despite their diversity, such signals share a common challenge: information is not uniformly distributed over time. Instead, temporal data often exhibits a mixture of smooth trends, persistent structures, and specific events that carry disproportionate importance \cite{cheng2025comprehensive,sakib2025ensemble}. 

Over the years, a wide range of models have been proposed for time series analysis, including statistical models as well as modern deep learning architectures \cite{liu2021forecast,makridakis1994time}. Harvey et al. \cite{harvey1990estimation} and Han et al. \cite{han2019review} provided a comprehensive review of various time series prediction models that interested readers can take a look at. Traditional sequence modeling approaches attempt to capture these dynamics through increasingly expressive architectures. Autoregressive (AR), Moving Average (MA), and Autoregressive Moving Average (ARMA) models are such classical statistical approaches primarily used for statistical time series data \cite{ma2020hybrid,cox1981statistical}. The AR model predicts future values as a linear combination of past observations, the MA model captures dependencies on past error terms, and the ARMA model combines both to model temporal dependencies using past values and past noise \cite{mehdizadeh2020using}. However, real-world time series data are often non-stationary, exhibiting trend and seasonality \cite{cheng2015time}. To address this, researchers added an adaptive nature to the statistical models, for example, Autoregressive moving average (ARMA), where differencing is applied to transform non-stationary data into a stationary form. Despite this modification, such models still rely on linear assumptions and struggle to capture complex, nonlinear patterns and long-term dependencies present in modern time series data \cite{ma2020hybrid}.

\begin{figure}[t]
\centering
\includegraphics[width=\linewidth]{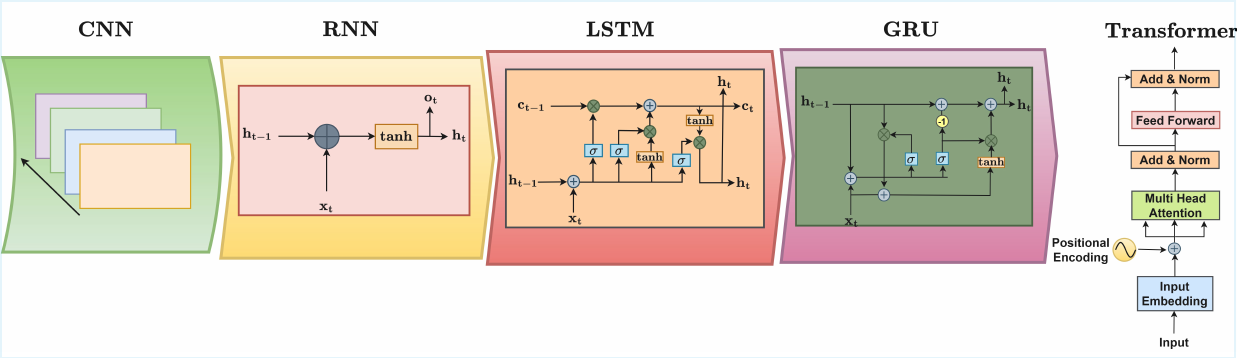}
\caption{Chronological evolution of time series modeling approaches. 
Time Series analysis methods followed by deep learning architectures, including CNN, RNN, LSTM, and GRU, culminating in Transformer-based models that leverage self-attention for capturing long-range dependencies.}
\label{fig:ts_evolution}
\end{figure}

Recent paradigms have shifted towards machine learning and deep learning architectures that can model complex, nonlinear relationships in time-series data \cite{park2026towards,ahmed2022review,chen2024graph}. Initial deep learning approaches applied convolutional neural networks (CNNs) to time series data, leveraging their ability to extract local patterns; however, CNNs lack an inherent mechanism to capture long-term temporal dependencies \cite{mo2025global}. To address this, recurrent neural networks (RNNs) were introduced, which model sequential data by maintaining hidden states over time \cite{sherstinsky2020fundamentals}. Nevertheless, RNNs suffer from issues such as vanishing and exploding gradients, making it difficult to learn long-range dependencies effectively \cite{ribeiro2020beyond}. Long short-term memory (LSTM) networks were then proposed to mitigate these issues through gated mechanisms that regulate information flow, but they are computationally expensive and still struggle with very long sequences \cite{sherstinsky2020fundamentals}. Gated recurrent units (GRUs) further simplified LSTMs by reducing the number of gates, improving efficiency; however, they still face limitations in capturing global context and parallelizing computations \cite{yunita2025performance}. In recent years, Transformer-based architectures have emerged as a powerful alternative, utilizing self-attention mechanisms to model long-range dependencies efficiently while enabling parallel processing, thereby addressing many of the limitations of earlier models \cite{zeng2023transformers,wang2025mamba}. As illustrated in Fig.~\ref{fig:ts_evolution}, time series models have evolved from statistical methods to increasingly complex deep learning architectures.

However, these advances come at the cost of increased architectural complexity, high computational demands, and limited interpretability, as most models function as black-box systems \cite{rudin2019stop}. In contrast, human perception of temporal signals is inherently selective and efficient: humans tend to emphasize salient variations, maintain compact representations of past information, and intuitively identify underlying trends \cite{pedrycz2014human}. This observation highlights the potential advantage of designing models that decompose time series into interpretable components, rather than relying exclusively on complex end-to-end sequence modeling \cite{10457027}. 

Motivated by the selective nature of human temporal perception, this work proposes \textbf{SPaRSe-TIME}, a decomposition-based framework that models time series as a combination of saliency, memory, and trend components, enabling selective emphasis on different temporal characteristics. To address the limitations of purely decomposition-based methods in complex multivariate settings, the framework is extended with interaction-aware embeddings and lightweight temporal mixing, allowing it to capture both cross-variable dependencies and local temporal dynamics while maintaining interpretability. Time series forecasting task is reformulated as an adaptive decomposition-learn-recombine strategy rather than learning everything explicitly. Extensive experiments across diverse datasets demonstrate that the proposed approach achieves competitive performance with significantly lower computational complexity and provides interpretable insights into how different components contribute to temporal reasoning. Moreover, this framework provides explicit insights into temporal reasoning by revealing how different components contribute to prediction across domains and makes the following contributions:
\begin{itemize}
\item \textbf{Decomposition-based temporal modeling:}
A novel framework, \textbf{SPaRSe-TIME}, is introduced to model time series as a structured combination of saliency, memory, and trend components, enabling selective and interpretable representation of temporal dynamics.

\item \textbf{Saliency-driven and low-rank formulation:}
Temporal modeling is formulated as a projection onto informative subspaces, where saliency acts as a sparsification operator and low-rank memory representations capture dominant temporal patterns efficiently.

\item \textbf{Computationally efficient architecture:}
A lightweight architecture is developed that avoids sequential recurrence and quadratic self-attention, achieving linear computational complexity while maintaining competitive performance across diverse datasets.

\item \textbf{Interpretable and adaptive temporal reasoning:}
The proposed framework provides explicit interpretability through learnable component weights, revealing how different temporal structures contribute to prediction and adapt across varying domains.
\end{itemize}

\section{Method}

\subsection{Problem Formulation}

Let $X \in \mathbb{R}^{T \times d}$ denote a multivariate time series of length $T$ with $d$ observed variables, and let $Y \in \mathbb{R}^{T \times m}$ denote the corresponding target signal. The objective is to learn a mapping
\begin{equation}
Y = f(X),
\end{equation}
where $f$ models the underlying temporal dependencies between input observations and target variables.

Conventional approaches treat $f$ as a sequence operator that processes all time steps uniformly, implicitly assuming that information is evenly distributed across the temporal dimension. However, real-world time series often exhibit heterogeneous structure, where informative patterns are sparsely distributed and interspersed with redundant or less relevant observations \cite{10858308}. In this work, it is hypothesized that effective temporal modeling can be achieved by selectively emphasizing informative components of the signal, rather than uniformly processing the entire sequence as echoed in early research works, for example, Liang et al. \cite{10.1108/02656711111109919}, and Oliveira et al. \cite{DEOLIVEIRA201627}. This perspective motivates a structured formulation in which temporal dynamics are captured through targeted representations of salient events, compressed historical context, and global trends.

\subsection{Overview of SPaRSe-TIME}

The proposed framework, \textbf{SPaRSe-TIME}, models temporal dynamics through a structured decomposition into three complementary components:
\begin{equation}
X \;\mapsto\; \big(S(X), \, M(X), \, G(X)\big),
\end{equation}
where $S(X)$, $M(X)$, and $G(X)$ correspond to saliency, memory, and trend representations, respectively. This decomposition induces a structured hypothesis space:
\begin{equation}
\hat{Y} = f\big(S(X), M(X), G(X)\big),
\end{equation}
which explicitly separates heterogeneous temporal information. Unlike conventional sequence models that process all time steps uniformly, the proposed formulation enables selective modeling of informative temporal components. The decomposition reflects a multi-scale temporal structure: $S(X)$ captures high-frequency variations, $M(X)$ encodes dominant low-dimensional patterns, and $G(X)$ models smooth long-range trends.

\subsection{Saliency-Projected Temporal Representation}

A saliency operator is introduced to identify informative time steps \cite{10.1145/3459637.3482446}. Let $w \in \mathbb{R}^{T}_{\geq 0}$ denote a normalized saliency vector defined as:
\begin{equation}
w_t = \frac{\|\nabla X_t\|_2}{\sum_{s=1}^{T} \|\nabla X_s\|_2}, 
\quad \text{where } \nabla X_t = X_t - X_{t-1}.
\end{equation}

Let $D_w \in \mathbb{R}^{T \times T}$ be a diagonal matrix with $(D_w)_{tt} = w_t$. The saliency-projected signal is then given by:
\begin{equation}
S(X) = D_w X.
\end{equation}

This operation can be interpreted as a projection onto a subspace spanned by informative temporal directions:
\begin{equation}
S(X) = \Pi_{\mathcal{S}}(X), \quad \mathcal{S} = \mathrm{span}\{e_t : w_t > 0\},
\end{equation}
where $\{e_t\}$ denotes the canonical basis. Unlike attention mechanisms, this projection is sparse, localized, and computationally efficient.

\subsection{Low-Rank Memory Representation}

Temporal sequences often exhibit redundancy and lie near a low-dimensional manifold \cite{gupta2024low,belay2024multivariate}. This property is exploited through a rank-$k$ approximation:
\begin{equation}
X \approx X_k = U_k \Sigma_k V_k^\top,
\end{equation}
where $(U_k, \Sigma_k, V_k)$ denote the leading singular components with $k \ll \min(T, d)$. The memory representation is defined as a projection onto the principal subspace:
\begin{equation}
M(X) = U_k^\top X \in \mathbb{R}^{k \times d}.
\end{equation}

Equivalently, $M(X)$ solves the optimization problem:
\begin{equation}
M(X) = \arg\min_{Z \in \mathbb{R}^{k \times d}} \|X - U_k Z\|_F^2.
\end{equation}

This representation captures dominant temporal structures while suppressing noise, providing a compact summary of historical information.

\subsection{Global Trend Modeling}

To model coarse-scale temporal dynamics, a linear smoothing operator $P \in \mathbb{R}^{T \times T}$ is introduced:
\begin{equation}
G(X) = P X,
\end{equation}
where $P$ acts as a low-pass filter (e.g., moving average or pooling operator). This induces a decomposition:
\begin{equation}
X = X_{\text{low}} + X_{\text{high}}, \quad
X_{\text{low}} = G(X), \quad
X_{\text{high}} = (I - P)X,
\end{equation}
separating low-frequency trends from high-frequency variations. The trend component $G(X)$ captures long-term temporal structure.

\subsection{Structured Temporal Mapping}

The three components are integrated through a weighted composition:
\begin{equation}
H = \sum_{i \in \{s,m,g\}} \alpha_i \, \phi_i(Z_i),
\quad Z_s = S(X), \; Z_m = M(X), \; Z_g = G(X),
\end{equation}
where $\phi_i$ are learnable linear transformations:
\begin{equation}
\phi_i(Z) = W_i Z + b_i.
\end{equation}

The weights $\alpha_i$ are parameterized using a softmax function:
\begin{equation}
\alpha_i = \frac{\exp(e_i)}{\sum_{j} \exp(e_j)}, 
\quad \alpha_i \geq 0, \quad \sum_i \alpha_i = 1.
\end{equation}

The final prediction is obtained as:
\begin{equation}
\hat{Y} = \psi(H),
\end{equation}
where $\psi$ is a nonlinear mapping.

The proposed formulation interprets temporal modeling as a composition of structured projections:
\begin{itemize}
\item $S(X)$ performs a sparse projection onto salient temporal directions,
\item $M(X)$ projects the signal onto a low-rank subspace capturing dominant patterns,
\item $G(X)$ extracts low-frequency structure via smoothing.
\end{itemize}

This decomposition yields a computationally efficient model with linear complexity in $T$, while providing explicit interpretability through component-wise contributions. The framework enables disentangled and adaptive temporal reasoning, offering a principled alternative to monolithic sequence modeling approaches.

\subsection{Computational Complexity}

The computational cost of the proposed framework is dominated by a set of linear operations over the temporal dimension. Specifically, given an input $X \in \mathbb{R}^{T \times d}$:

\begin{itemize}
\item \textbf{Saliency computation:} The temporal gradient $\nabla X_t = X_t - X_{t-1}$ is computed for each time step, followed by normalization, resulting in a complexity of $\mathcal{O}(T d)$.

\item \textbf{Low-rank projection:} Projecting the signal onto a $k$-dimensional subspace requires $\mathcal{O}(k T d)$ operations, assuming precomputed basis vectors (or efficient truncated decomposition), where $k \ll \min(T,d)$.

\item \textbf{Linear transformations:} The mappings $\phi_i(Z) = W_i Z + b_i$ are applied independently to each component, yielding a total cost of $\mathcal{O}(T d)$.
\end{itemize}

Combining these terms, the overall complexity of the proposed method scales as:
\begin{equation}
\mathcal{O}(k T d),
\end{equation}
which is linear in the sequence length $T$.

In contrast, self-attention mechanisms require pairwise interactions across time steps \cite{pmlr-v201-duman-keles23a}, leading to $\mathcal{O}(T^2 d)$ complexity. Thus, for long sequences ($T \gg k$), the proposed framework achieves a substantial reduction in computational cost while maintaining expressive temporal modeling capacity. Table \ref{tab:complexity} describes a comparative complexity analysis of the proposed model with established models in existing literature.

\begin{table}[t]
\centering
\caption{Computational complexity comparison of different time series models. $T$ denotes sequence length, $d$ feature dimension, $k$ low-rank dimension, and $h$ hidden size.}
\label{tab:complexity}
\begin{tabular}{l c}
\toprule
\textbf{Model} & \textbf{Time Complexity} \\
\midrule
CNN & $\mathcal{O}(T d k)$ \\
RNN / LSTM & $\mathcal{O}(T h^2)$ \\
GRU & $\mathcal{O}(T h^2)$ \\
Transformer & $\mathcal{O}(T^2 d)$ \\
\midrule
\textbf{SPaRSe-TIME (Ours)} & $\mathbf{\mathcal{O}(k T d)}$ \\
\bottomrule
\end{tabular}
\end{table}

\section{Algorithm}

\begin{figure*}[t]
\centering
\includegraphics[width=\textwidth]{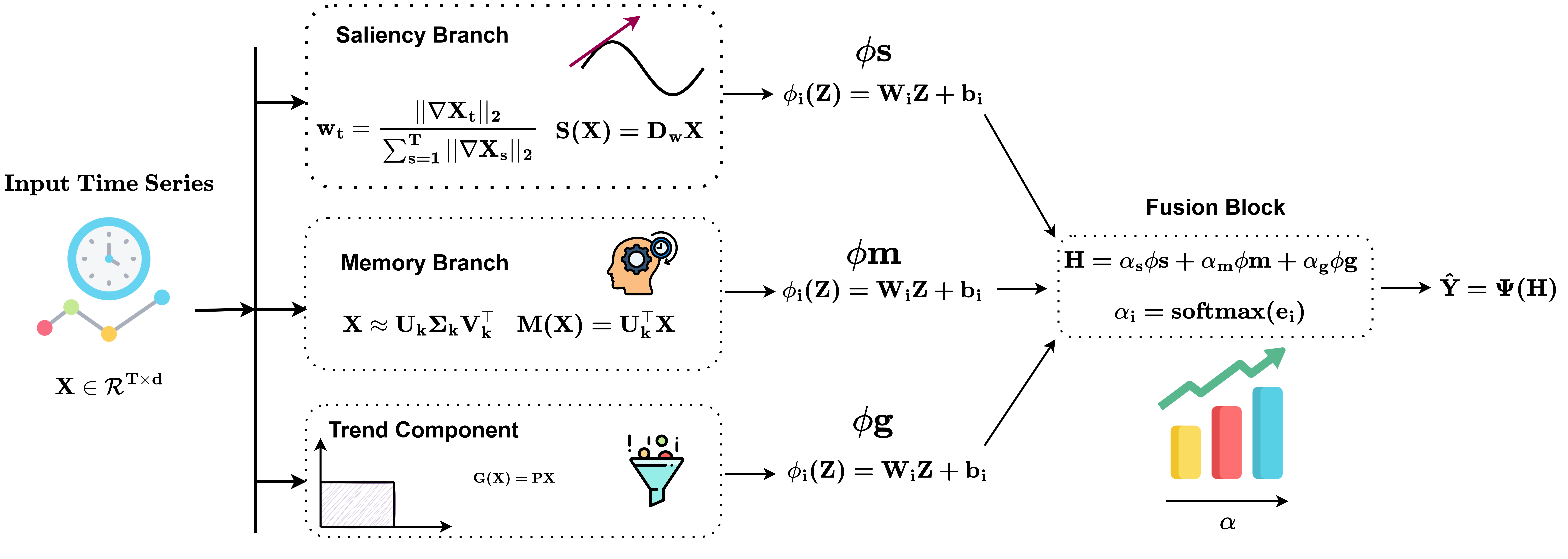}
\caption{
Overview of the proposed \textbf{SPaRSe-TIME} framework.}
\label{fig:sparse-time}
\end{figure*}

This section presents the computational procedure of the proposed \textbf{SPaRSe-TIME} framework. The model operates by decomposing the input time series into saliency, memory, and trend components, followed by a structured aggregation for prediction. Given an input time series $X \in \mathbb{R}^{T \times d}$, the algorithm first computes a saliency representation by estimating temporal gradients and assigning normalized importance weights to each time step. This yields a weighted projection $S(X)$ that emphasizes informative temporal variations. Next, a low-rank memory representation is constructed by projecting the input onto a $k$-dimensional subspace obtained via truncated singular value decomposition (SVD). This step captures dominant temporal patterns while reducing redundancy. To model long-term dynamics, a smoothing operator is applied to obtain the trend component $G(X)$, which isolates low-frequency temporal structure. The three components are then independently transformed through learnable linear mappings and combined using adaptive weights constrained on a simplex. The aggregated representation is finally passed through a nonlinear function to produce the prediction. The overall procedure is summarized in Algorithm~\ref{alg:sparse-time}. Fig.~\ref{fig:sparse-time} illustrates the proposed SPaRSe-TIME framework.

\begin{algorithm}[t]
\caption{SPaRSe-TIME: Saliency-Projected Low-Rank Temporal Modeling}
\label{alg:sparse-time}
\begin{algorithmic}[1]
\REQUIRE Time series $X \in \mathbb{R}^{T \times d}$, rank $k$
\ENSURE Prediction $\hat{Y}$

\STATE \textbf{// Saliency Computation}
\FOR{$t = 2$ to $T$}
    \STATE $\nabla X_t \leftarrow X_t - X_{t-1}$
\ENDFOR
\STATE $w_t \leftarrow \|\nabla X_t\|_2$
\STATE Normalize $w \leftarrow w / \sum_{t=1}^{T} w_t$
\STATE $D_w \leftarrow \mathrm{diag}(w_1, \dots, w_T)$
\STATE $S(X) \leftarrow D_w X$

\vspace{0.5em}
\STATE \textbf{// Low-Rank Memory Projection}
\STATE Compute top-$k$ SVD: $X \approx U_k \Sigma_k V_k^\top$
\STATE $M(X) \leftarrow U_k^\top X$

\vspace{0.5em}
\STATE \textbf{// Global Trend Extraction}
\STATE Apply smoothing operator $P$
\STATE $G(X) \leftarrow P X$

\vspace{0.5em}
\STATE \textbf{// Component-wise Transformation}
\STATE $Z_s \leftarrow \phi_s(S(X))$
\STATE $Z_m \leftarrow \phi_m(M(X))$
\STATE $Z_g \leftarrow \phi_g(G(X))$

\vspace{0.5em}
\STATE \textbf{// Adaptive Weighting}
\STATE $\alpha_i \leftarrow \exp(e_i) / \sum_j \exp(e_j), \quad i \in \{s,m,g\}$

\vspace{0.5em}
\STATE \textbf{// Aggregation}
\STATE $H \leftarrow \alpha_s Z_s + \alpha_m Z_m + \alpha_g Z_g$

\vspace{0.5em}
\STATE \textbf{// Prediction}
\STATE $\hat{Y} \leftarrow \psi(H)$

\RETURN $\hat{Y}$
\end{algorithmic}
\end{algorithm}

\section{Experimental Setup}
\subsection{Dataset Description}

To evaluate the generality and robustness of the proposed \textit{SPaRSe-TIME} framework, experiments are conducted on five diverse real-world time series datasets spanning energy consumption, financial markets, environmental sensing, and atmospheric conditions. These datasets exhibit varying temporal characteristics, including high-frequency volatility, seasonal trends, and long-term dependencies. 

\subsubsection{Individual Household Electric Power Consumption \cite{individual_household_electric_power_consumption_235}}

This dataset contains minute-level measurements of electric power consumption for a single household over multiple years(December 2006 - November 2010). It includes features such as global active power, voltage, and sub-metering variables. The data exhibits strong daily and weekly seasonality, along with occasional spikes due to appliance usage, making it suitable for evaluating both short-term memory and long-term trend modeling.

\subsubsection{Netflix Stock Price Dataset\cite{netflix_kaggle}}

The Netflix stock dataset consists of daily historical price data, including open, high, low, close, adjusted close, and trading volume. Financial time series are inherently noisy and exhibit non-stationary behavior with abrupt fluctuations driven by market dynamics. This dataset is used to assess the model’s ability to capture high-frequency variations (saliency) and temporal dependencies (memory).

\subsubsection{NASDAQ-100 Stock Price Dataset \cite{nasdaq_kaggle}}

The NASDAQ-100 dataset contains historical stock price data for multiple companies over an extended time period. Similar to the Netflix dataset, it includes standard financial indicators such as open, high, low, close, and volume. The dataset introduces additional complexity due to cross-company variability and diverse temporal patterns, making it suitable for evaluating model generalization across different financial instruments.

\subsubsection{Weather Dataset \cite{weather_kaggle}}

The weather dataset contains meteorological observations such as temperature, humidity, wind speed, pressure, and visibility. Unlike financial data, weather signals are relatively smooth and exhibit strong seasonal and low-frequency patterns. This dataset is particularly useful for evaluating the trend component of the model, as long-term temporal dynamics dominate predictive performance.

\subsubsection{UCI Air Quality Dataset \cite{air_quality_360}}

The UCI Air Quality dataset consists of hourly averaged responses from chemical sensors measuring air pollutants, along with meteorological variables such as temperature and humidity. The data includes both periodic patterns and sensor noise, presenting a challenging scenario where both memory and trend components are important. It provides a balanced benchmark for evaluating the interplay between different temporal components.

\subsection{Data Preprocessing}

A unified preprocessing pipeline consisting of normalization, sliding window construction, and component decomposition is adopted to ensure consistency across heterogeneous datasets. For each dataset, a subset of relevant numerical features is selected based on the context of the dataset. 

\subsubsection{Normalization}

First, to stabilize optimization and ensure comparable feature scales, z-score normalization is applied:
\[
\tilde{x}_t = \frac{x_t - \mu}{\sigma + \epsilon},
\]
where $\mu$ and $\sigma$ denote the mean and standard deviation computed over the training set, and $\epsilon$ is a small constant for numerical stability. This transformation ensures:
\[
\mathbb{E}[\tilde{x}] = 0, \quad \mathrm{Var}(\tilde{x}) = 1.
\]

\subsubsection{Sliding Window Construction}

Then, a sliding window approach is employed to convert time series data into supervised learning samples. For a window size $L$, each sample is constructed as:
\[
X^{(i)} = \{x_i, x_{i+1}, \dots, x_{i+L-1}\}, \quad y^{(i)} = x_{i+L},
\]
where $X^{(i)} \in \mathbb{R}^{L \times d}$ and $y^{(i)}$ is the prediction target. This formulation allows the model to capture temporal dependencies within a fixed-length context.

\subsubsection{Temporal Decomposition}

Finally, each input sequence was decomposed into three interpretable components: saliency, memory, and trend.

\paragraph{Saliency (High-Frequency Component)}
The saliency component captures local fluctuations and abrupt changes:
\[
S_t = \|x_t - x_{t-1}\|,
\]
where $\|\cdot\|$ denotes the element-wise absolute difference.

\paragraph{Memory (Local Context)}
The memory component corresponds to the original signal:
\[
M_t = x_t,
\]
which preserves short-term temporal dependencies.

\paragraph{Trend (Low-Frequency Component)}
To extract long-term dynamics, a moving average was computed:
\[
G_t = \frac{1}{k} \sum_{i=t-\lfloor k/2 \rfloor}^{t+\lfloor k/2 \rfloor} x_i,
\]
where $k$ is the smoothing window size. This operation suppresses high-frequency noise and reveals the underlying trend. Each training instance is represented as:
\[
(S^{(i)}, M^{(i)}, G^{(i)}),
\]
where each component lies in $\mathbb{R}^{L \times d}$. This structured representation enables the model to learn disentangled temporal patterns at multiple scales.

\subsubsection{Data Splitting}

A chronological split of each dataset into training (70\%), validation (15\%), and test (15\%) sets was performed to prevent data leakage. All normalization statistics ($\mu, \sigma$) are computed exclusively on the training set and applied to the remaining splits.

\subsection{Model Architecture}

\textit{SPaRSe-TIME}, a lightweight and interpretable framework, is proposed that models temporal dynamics by decomposing them into saliency, memory, and trend components.

\subsubsection{Component Projection}

Given the decomposed inputs $(S, M, G)$, where each component lies in $\mathbb{R}^{L \times d}$, they are first projected into a shared latent space:
\[
h_s = W_s S, \quad h_m = W_m M, \quad h_g = W_g G,
\]
where $W_s, W_m, W_g \in \mathbb{R}^{d \times h}$ are learnable linear transformations and $h$ is the hidden dimension. These projections allow the model to learn distinct representations for high-frequency (saliency), short-term (memory), and low-frequency (trend) components.

\subsubsection{Adaptive Component Weighting}

To dynamically balance the contribution of each component, a learnable weighting mechanism is introduced. Let $\theta \in \mathbb{R}^3$ denote the parameter vector. The normalized weights are computed using a softmax function:
\[
\alpha = \text{softmax}(\theta), \quad \sum_{i=1}^{3} \alpha_i = 1.
\]

The final representation is obtained as a weighted combination:
\[
H = \alpha_1 h_s + \alpha_2 h_m + \alpha_3 h_g.
\]

This formulation enables the model to adaptively emphasize different temporal components depending on the dataset characteristics.

\subsubsection{Nonlinear Transformation and Prediction}

The combined representation $H$ is passed through a nonlinear activation function and a fully connected output layer:
\[
\hat{y} = f(H) = W_o \, \sigma(H),
\]
where $\sigma(\cdot)$ denotes the ReLU activation function and $W_o \in \mathbb{R}^{h \times 1}$ is a learnable projection. The model outputs a prediction $\hat{y}$ for each time step in the sequence.

\subsubsection{Interpretability}

A key advantage of the proposed architecture is its interpretability. The learned weights $\alpha$ provide a direct measure of the relative importance of saliency, memory, and trend components:
\[
\alpha = (\alpha_S, \alpha_M, \alpha_G).
\]
This allows us to analyze how different temporal structures contribute to prediction across datasets, offering insights into domain-specific temporal dynamics.

\subsection{Training Setup}

The model is trained using the \texttt{AdamW} optimizer, which adapts learning rates for each parameter based on first and second-order moments of the gradients. Given model parameters $\Theta$, the update rule follows:
\[
\Theta \leftarrow \Theta - \eta \, \nabla_{\Theta} \mathcal{L},
\]
where $\eta$ is the learning rate. In all experiments, we set $\eta = 10^{-3}$. The problem is formulated as a regression task and minimizes the Mean Squared Error (MSE) loss:
\[
\mathcal{L} = \frac{1}{N} \sum_{i=1}^{N} \left( y^{(i)} - \hat{y}^{(i)} \right)^2,
\]
where $y^{(i)}$ and $\hat{y}^{(i)}$ denote the ground truth and predicted values, respectively, and $N$ is the number of samples. To prevent overfitting, weight decay (L2 regularization) was employed during optimization. The objective becomes:
\[
\mathcal{L}_{\text{total}} = \mathcal{L} + \lambda \|\Theta\|_2^2,
\]
where $\lambda$ is the regularization coefficient. Early stopping based on validation loss was also adopted to improve generalization. Training was terminated if the validation loss did not improve for a fixed number of epochs (patience). The model parameters corresponding to the best validation performance were retained for testing. Training was performed using mini-batches of size $B$. Each batch consisted of triplets $(S, M, G)$ along with corresponding targets. The model was trained for a maximum of 50-100 epochs, depending on the dataset, with early stopping applied to avoid overfitting. All models were implemented in PyTorch and trained on a GPU-enabled environment. For fair comparison, all baseline models are trained under the same preprocessing pipeline and experimental settings.

\subsection{Evaluation Metrics}

To comprehensively evaluate model performance, multiple standard regression metrics were employed that capture different aspects of prediction accuracy.

\paragraph{Mean Absolute Error (MAE)}
MAE measures the average absolute difference between predicted and true values:
\[
\text{MAE} = \frac{1}{N} \sum_{i=1}^{N} \left| y^{(i)} - \hat{y}^{(i)} \right|.
\]

\paragraph{Root Mean Squared Error (RMSE)}
RMSE penalizes larger errors more heavily and is defined as:
\[
\text{RMSE} = \sqrt{ \frac{1}{N} \sum_{i=1}^{N} \left( y^{(i)} - \hat{y}^{(i)} \right)^2 }.
\]
\paragraph{Coefficient of Determination ($R^2$)}
The $R^2$ score evaluates how well the model explains the variance in the data:
\[
R^2 = 1 - \frac{\sum_{i=1}^{N} (y^{(i)} - \hat{y}^{(i)})^2}{\sum_{i=1}^{N} (y^{(i)} - \bar{y})^2},
\]
where $\bar{y}$ is the mean of the ground truth values. These metrics provide complementary perspectives, with MAE capturing average error, RMSE emphasizing large deviations, and $R^2$ measuring goodness of fit.

\subsubsection{Baselines for Comparison}

To assess the effectiveness of the proposed SPaRSe-TIME framework, it was compared against several widely used time series forecasting models.

\paragraph{Recurrent Neural Network (RNN)}
The vanilla RNN models sequential dependencies by maintaining a hidden state updated at each time step:
\[
h_t = \sigma(W_h h_{t-1} + W_x x_t),
\]
where $h_t$ is the hidden state and $\sigma(\cdot)$ is a nonlinear activation function. While simple, RNNs often struggle with long-term dependencies.

\paragraph{Gated Recurrent Unit (GRU)}
GRU introduces gating mechanisms to control information flow and mitigate vanishing gradients. It uses update and reset gates to adaptively retain or discard past information, enabling better modeling of temporal dependencies.

\paragraph{Long Short-Term Memory (LSTM)}
LSTM further extends RNNs with memory cells and gating mechanisms (input, forget, and output gates) to capture long-range dependencies:
\[
c_t = f_t \odot c_{t-1} + i_t \odot \tilde{c}_t,
\]
where $c_t$ denotes the cell state and $\odot$ represents element-wise multiplication.

\paragraph{Transformer}
The Transformer model leverages self-attention to model global dependencies:
\[
\text{Attention}(Q, K, V) = \text{softmax}\left( \frac{QK^T}{\sqrt{d}} \right)V,
\]
allowing the model to capture relationships across all time steps simultaneously without recurrence.

\section{Results}

\subsection{Quantitative Results on Household Dataset \cite{individual_household_electric_power_consumption_235}}

Table~\ref{tab:household_results} presents the performance comparison of all models on the Individual Household Electric Power Consumption dataset.

\begin{table}[t]
\centering
\caption{Performance comparison on Household dataset}
\label{tab:household_results}
\begin{tabular}{lcccc}
\hline
Model & MAE $\downarrow$ & RMSE $\downarrow$ & $R^2$ $\uparrow$ \\
\hline
RNN          & 0.458 & 0.621 & 0.375 \\
GRU          & 0.467 & 0.642 & 0.334 \\
LSTM         & 0.457 & 0.633 & 0.353 \\
Transformer  & 0.439 & 0.616 & 0.385 \\
\textbf{SPaRSe-TIME}  & \textbf{0.398} & \textbf{0.554} & \textbf{0.503} \\
\hline
\end{tabular}
\end{table}

The proposed SPaRSe-TIME model consistently outperforms all baseline models across all evaluation metrics. In particular, it achieves the lowest MAE and RMSE, along with the highest $R^2$ score, indicating superior predictive accuracy and better variance explanation. Compared to the Transformer, which is the strongest baseline, SPaRSe-TIME reduces RMSE significantly while using substantially fewer computational resources.

\begin{figure}[h]
\centering
\includegraphics[width=0.9\linewidth]{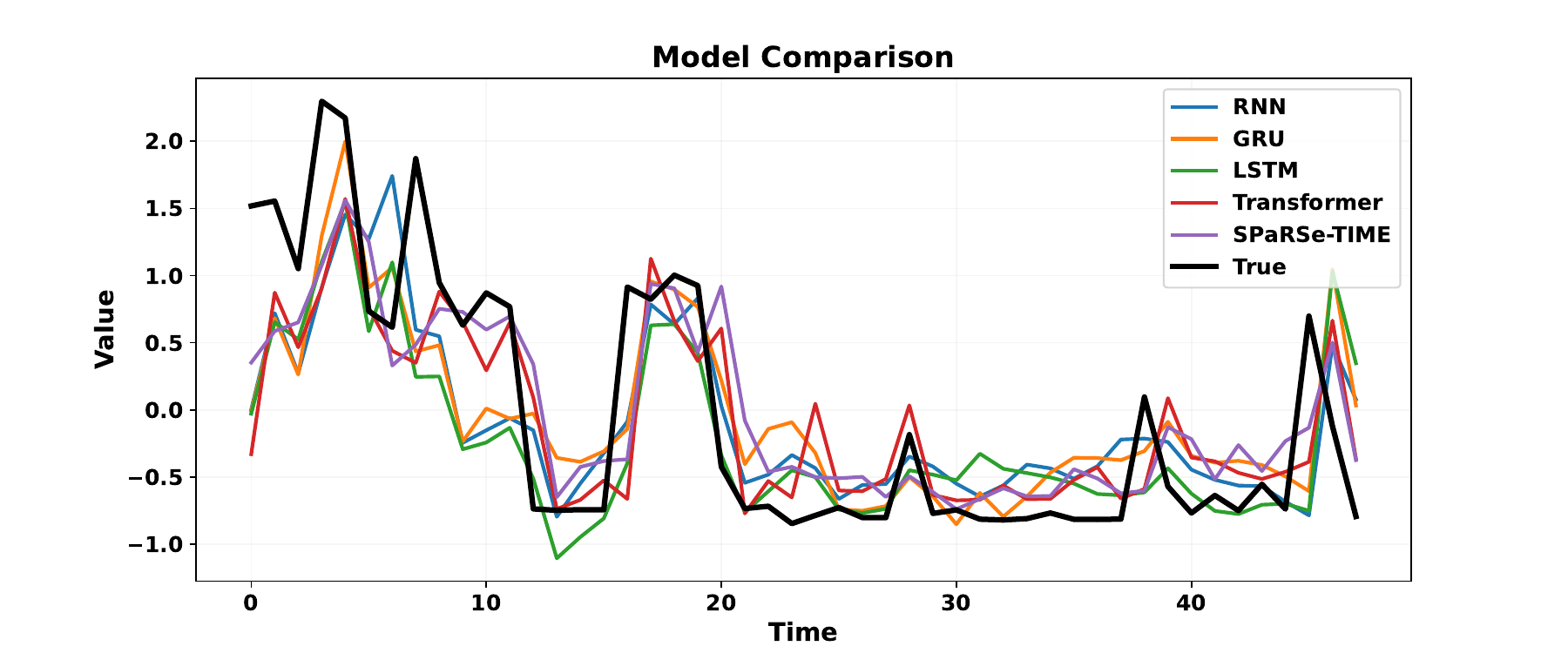}
\caption{Prediction comparison across different models. SPaRSe-TIME produces smoother and more accurate predictions compared to recurrent and attention-based baselines.}
\label{fig:model_comparison}
\end{figure}

Figure~\ref{fig:model_comparison} shows the prediction trajectories of different models. Traditional RNN-based models exhibit higher variance and struggle to capture stable patterns, while the Transformer captures global dependencies but remains noisy. In contrast, SPaRSe-TIME produces smoother and more consistent predictions, closely aligning with the ground truth.

\begin{figure}[h]
\centering
\includegraphics[width=\linewidth]{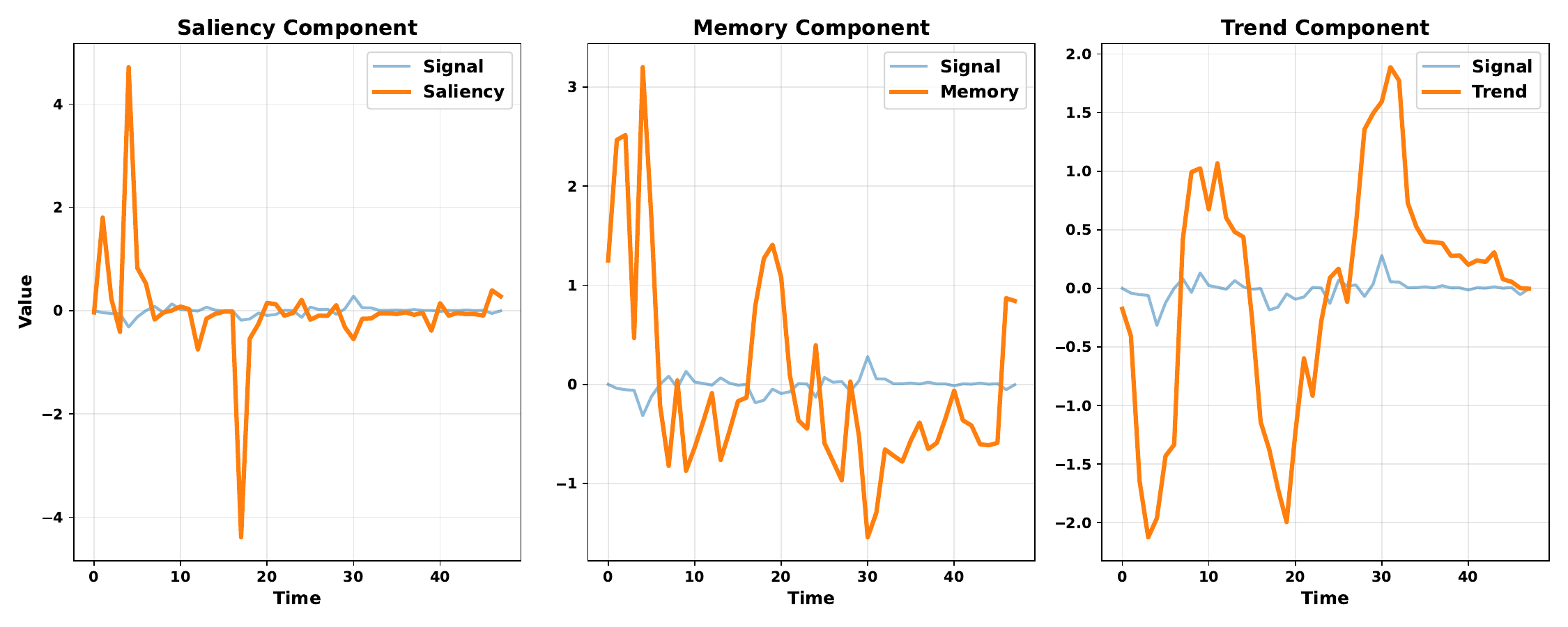}
\caption{Decomposition of the input signal into saliency, memory, and trend components. Each component captures distinct temporal characteristics.}
\label{fig:components}
\end{figure}

Figure~\ref{fig:components} illustrates the learned decomposition. The saliency component captures high-frequency fluctuations and abrupt changes, the memory component models local temporal dependencies, and the trend component reflects smooth, low-frequency dynamics. This disentanglement allows the model to separately learn different temporal structures.

\begin{figure}[h]
\centering
\includegraphics[width=0.9\linewidth]{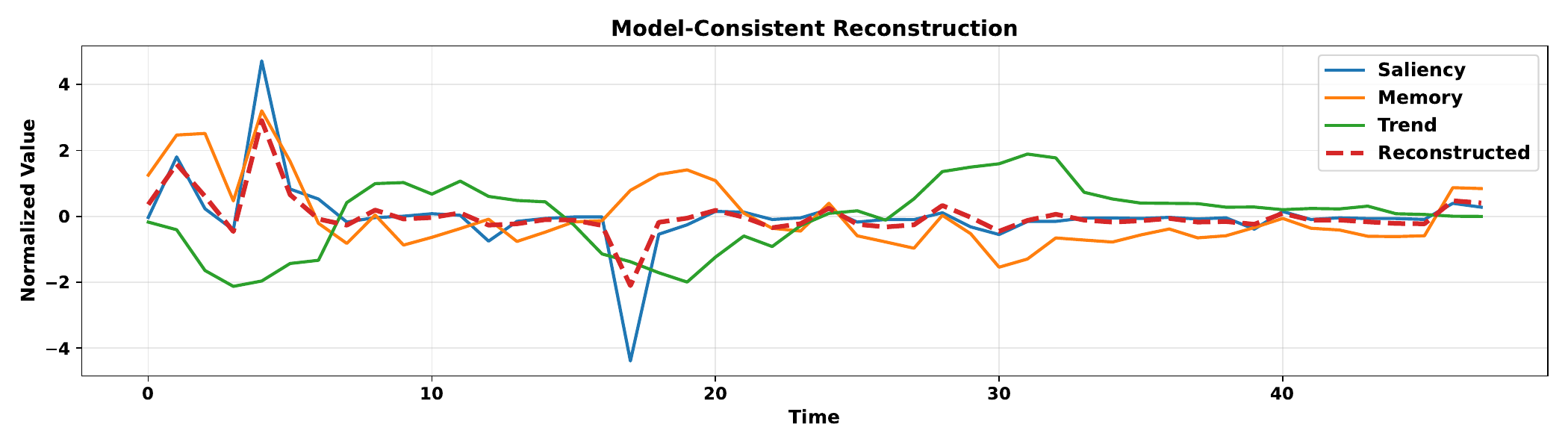}
\caption{Reconstruction of the signal using a learned weighted combination of saliency, memory, and trend components.}
\label{fig:reconstruction}
\end{figure}

Figure~\ref{fig:reconstruction} demonstrates that the original signal can be effectively reconstructed as a learned weighted combination of the three components. The reconstructed signal closely follows the true dynamics, validating the effectiveness of the decomposition. This also highlights the interpretability of the model, as each component contributes meaningfully to the final prediction.

\subsection{Results on Weather Dataset \cite{weather_kaggle}}

\subsubsection{Quantitative Results}

Table~\ref{tab:weather_results} presents the performance comparison across all models on the weather dataset. SPaRSe-TIME achieves the best performance across all evaluation metrics, demonstrating its effectiveness in modeling complex temporal dynamics. Compared to the strongest baseline (Transformer), the proposed model achieves lower MAE and RMSE while also improving the $R^2$ score, indicating better predictive accuracy and variance explanation.

\begin{table}[t]
\centering
\caption{Performance comparison on Weather dataset}
\label{tab:weather_results}
\begin{tabular}{lccc}
\hline
Model & MAE $\downarrow$ & RMSE $\downarrow$ & $R^2$ $\uparrow$ \\
\hline
RNN          & 0.548 & 0.704 & 0.471 \\
GRU          & 0.568 & 0.728 & 0.433 \\
LSTM         & 0.572 & 0.731 & 0.428 \\
Transformer  & 0.518 & 0.668 & 0.524 \\
\textbf{SPaRSe-TIME}  & \textbf{0.475} & \textbf{0.635} & \textbf{0.568} \\
\hline
\end{tabular}
\end{table}

\begin{figure}[h]
\centering
\includegraphics[width=\linewidth]{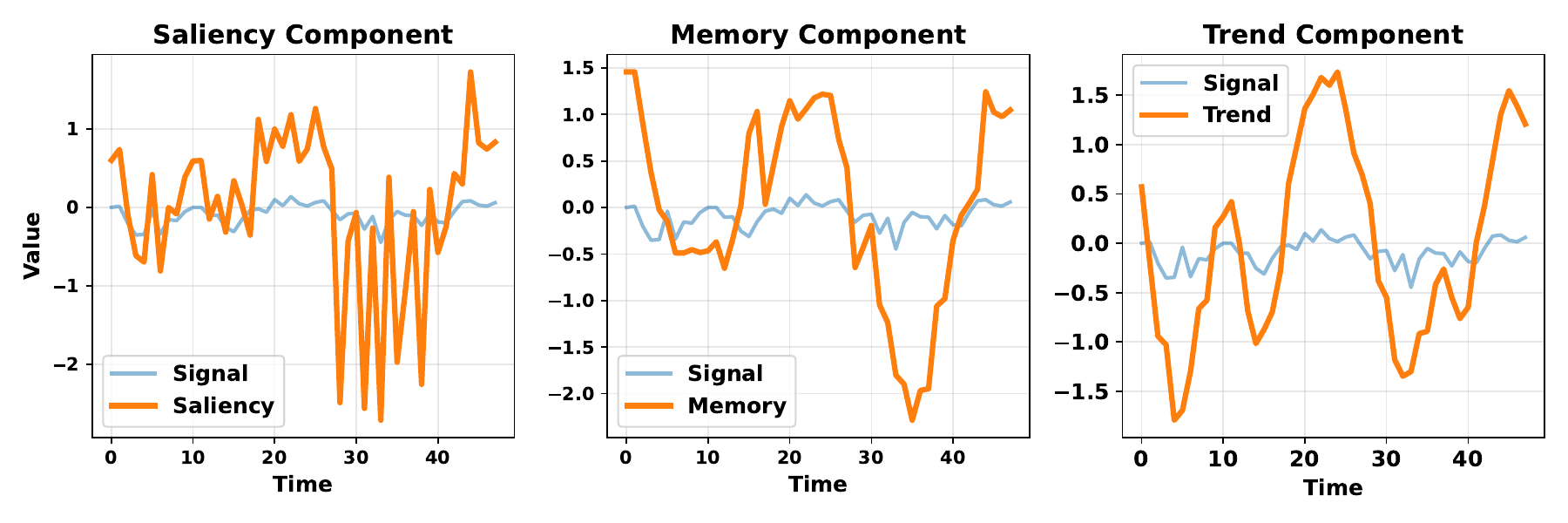}
\caption{Decomposition into saliency, memory, and trend components on the weather dataset.}
\label{fig:weather_components}
\end{figure}

Figure~\ref{fig:weather_components} illustrates the decomposition of the signal. The saliency component captures rapid fluctuations, the memory component models short-term temporal dependencies, and the trend component captures smooth long-term variations. Unlike other datasets, the trend component plays a more significant role, reflecting the seasonal and gradual changes in weather patterns.

\begin{figure}[t]
\centering
\includegraphics[width=0.9\linewidth]{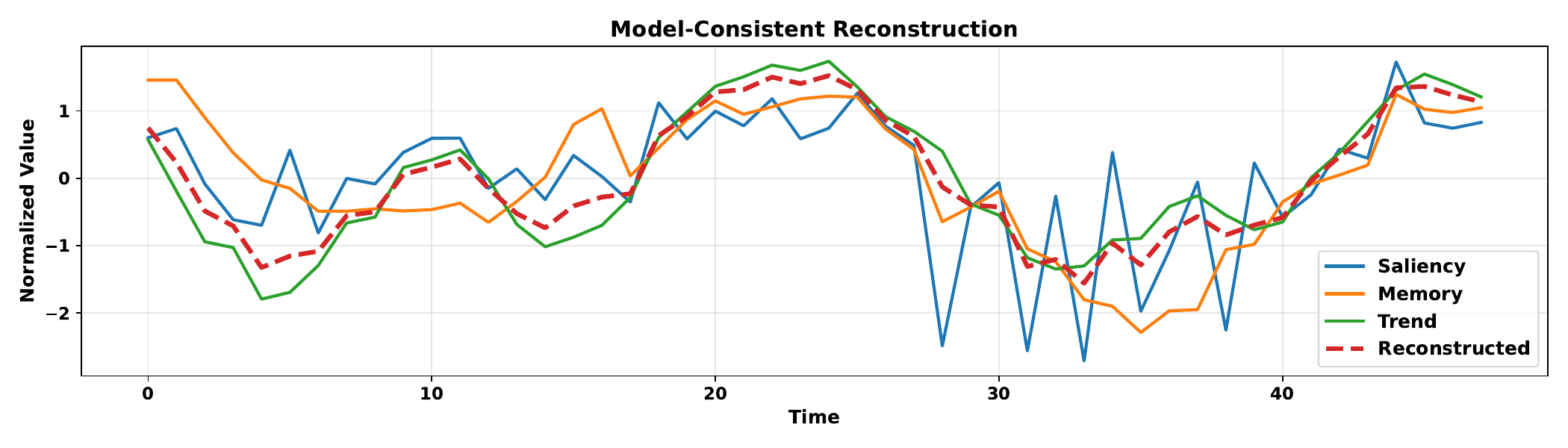}
\caption{Reconstruction from saliency, memory, and trend components.}
\label{fig:weather_reconstruction}
\end{figure}

Figure~\ref{fig:weather_reconstruction} shows that the original signal can also be effectively reconstructed using the weighted combination of the three components. The reconstructed signal closely follows the underlying temporal structure, validating the effectiveness of the decomposition.

\subsection{Results on Netflix Dataset \cite{netflix_kaggle}}

\subsubsection{Quantitative Results}

Table~\ref{tab:netflix_results} presents the performance comparison across all models on the Netflix stock price dataset. Unlike other datasets, all models achieve relatively similar performance on the Netflix dataset, with very low $R^2$ scores close to zero. This indicates that the dataset exhibits highly noisy and less predictable behavior, making it challenging for all models to capture meaningful temporal patterns. SPaRSe-TIME achieves competitive performance, obtaining the lowest RMSE and highest $R^2$, although the overall improvement remains marginal due to the inherent difficulty of the task.

\begin{table}[t]
\centering
\caption{Performance comparison on Netflix dataset}
\label{tab:netflix_results}
\begin{tabular}{lccc}
\hline
Model & MAE $\downarrow$ & RMSE $\downarrow$ & $R^2$ $\uparrow$ \\
\hline
RNN          & 0.518 & 0.842 & 0.010 \\
GRU          & 0.520 & 0.845 & 0.004 \\
LSTM         & \textbf{0.513} & 0.842 & 0.010 \\
Transformer  & 0.563 & 0.892 & -0.110 \\
\textbf{SPaRSe-TIME}  & 0.522 & \textbf{0.841} & \textbf{0.013} \\
\hline
\end{tabular}
\end{table}

\begin{figure}[h]
\centering
\includegraphics[width=\linewidth]{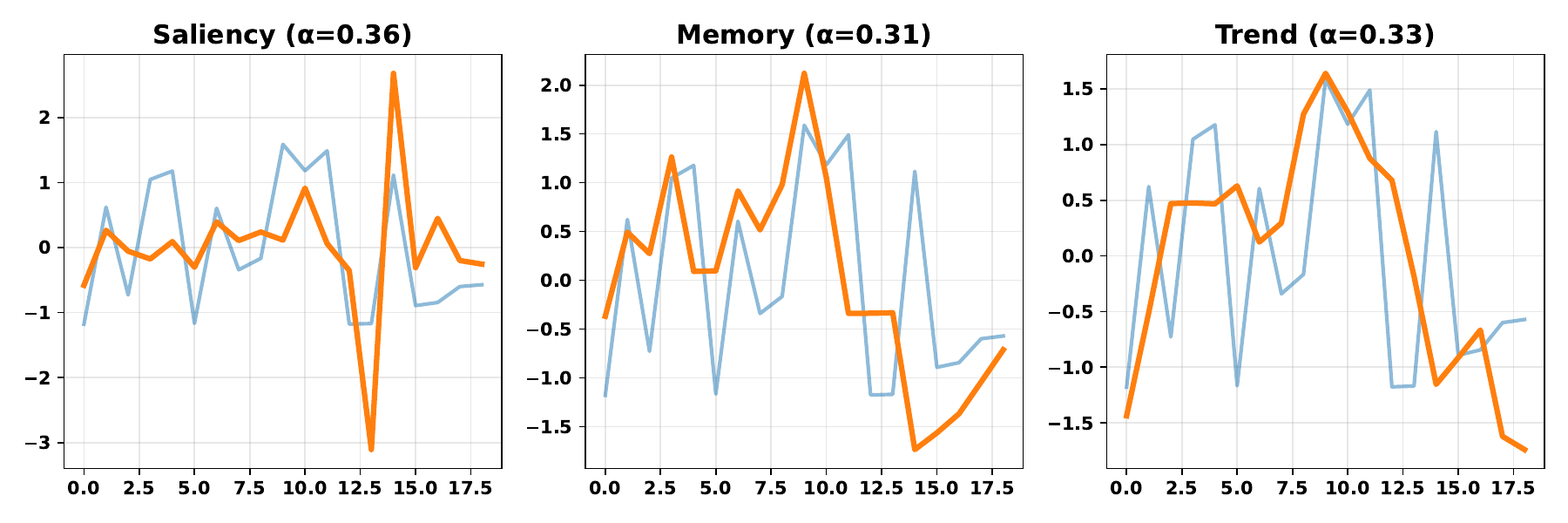}
\caption{Saliency, memory, and trend decomposition on the Netflix dataset.}
\label{fig:netflix_components}
\end{figure}

Figure~\ref{fig:netflix_components} shows the decomposition of the signal. The saliency component captures high-frequency fluctuations, while the memory component reflects short-term dependencies. The trend component is relatively weak and unstable compared to other datasets, indicating the absence of strong long-term patterns in stock price movements.

\begin{figure}[h]
\centering
\includegraphics[width=0.9\linewidth]{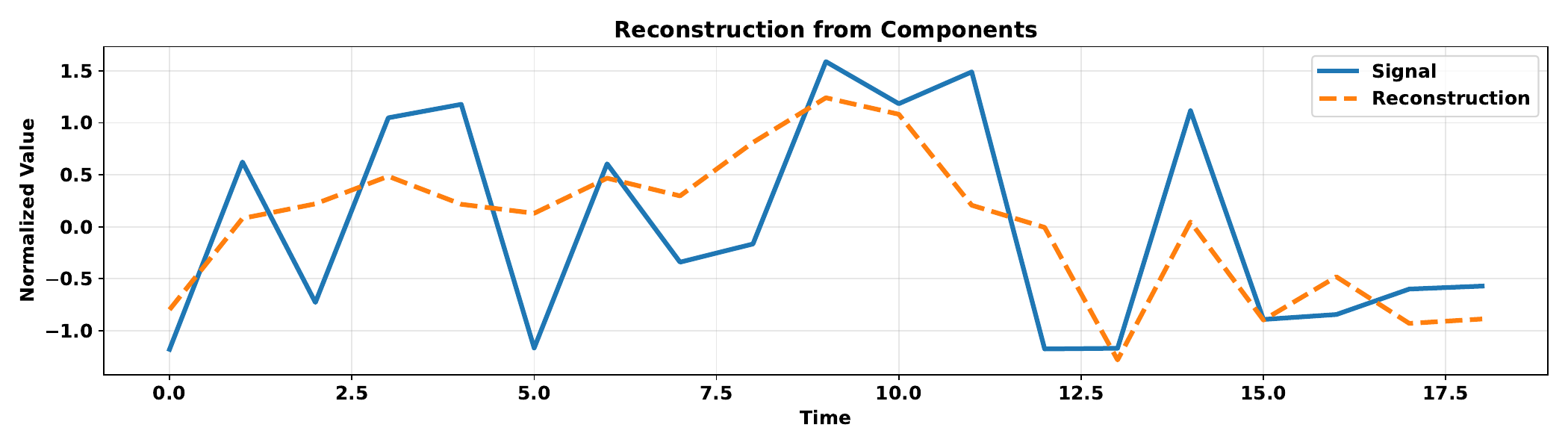}
\caption{Reconstruction from decomposed components on the Netflix dataset.}
\label{fig:netflix_reconstruction}
\end{figure}

Figure~\ref{fig:netflix_reconstruction} illustrates that the reconstructed signal follows the general structure of the original signal but fails to capture sharp fluctuations accurately. This highlights the difficulty of modeling stock price dynamics, where noise and sudden changes dominate the signal.

\subsection{Results on NASDAQ Dataset \cite{nasdaq_kaggle}}

Table~\ref{tab:nasdaq_results} presents the performance comparison across all models on the NASDAQ-100 dataset.

\begin{table}[h]
\centering
\caption{Performance comparison on NASDAQ dataset}
\label{tab:nasdaq_results}
\begin{tabular}{lccc}
\hline
Model & MAE $\downarrow$ & RMSE $\downarrow$ & $R^2$ $\uparrow$ \\
\hline
Naive        & 0.0240 & 0.0338 & -0.8376 \\
RNN          & 0.0178 & 0.0254 & -0.0378 \\
GRU          & 0.0176 & 0.0254 & -0.0398 \\
LSTM         & 0.0174 & 0.0252 & -0.0211 \\
Transformer  & \textbf{0.0173} & \textbf{0.0250} & \textbf{-0.0070} \\
SPaRSe-TIME  & 0.0193 & 0.0267 & -0.1476 \\
\hline
\end{tabular}
\end{table}

All models achieve very similar performance on the NASDAQ dataset, with $R^2$ values close to zero or negative. This indicates that the dataset is highly noisy and difficult to predict, with weak temporal structure. The Transformer achieves the best overall performance, while SPaRSe-TIME underperforms compared to other models in this setting.

\begin{figure}[h]
\centering
\includegraphics[width=\linewidth]{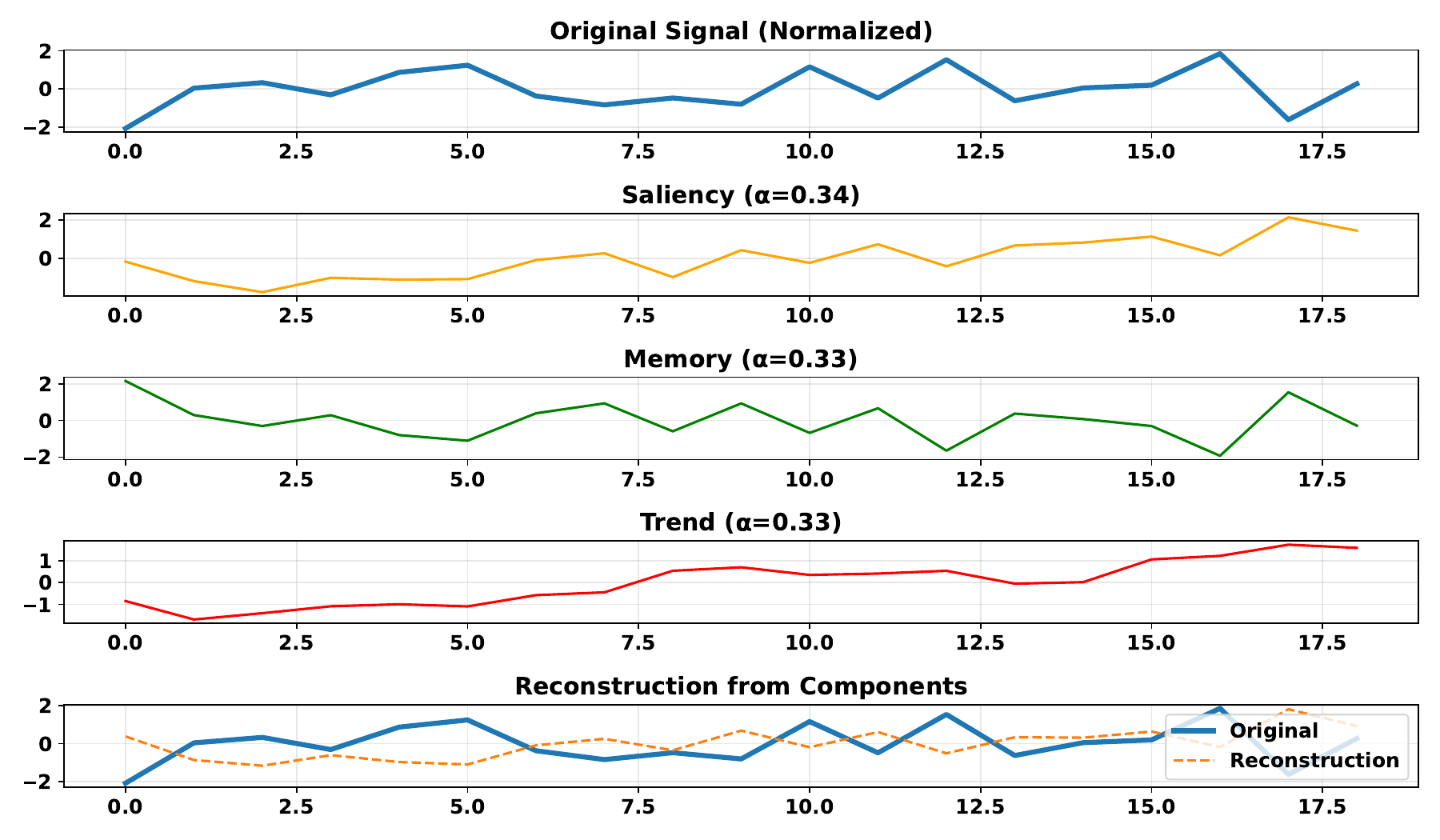}
\caption{Decomposition of NASDAQ signal into saliency, memory, and trend components.}
\label{fig:nasdaq_components}
\end{figure}

Figure~\ref{fig:nasdaq_components} illustrates the decomposition of the NASDAQ signal. The saliency and memory components dominate the representation, capturing short-term fluctuations and local dependencies. The trend component is relatively weak and smooth, indicating the absence of strong long-term patterns in financial return series. The reconstruction from saliency, memory, and trend components captures the general direction of the signal but struggles to follow sharp fluctuations accurately. This reflects the inherently stochastic nature of financial return data, where noise dominates structured temporal patterns.

The NASDAQ dataset highlights a key limitation of temporal decomposition approaches in highly noisy environments. Unlike weather or household datasets, financial return series exhibit a low signal-to-noise ratio and minimal long-term structure. As a result, all models, including SPaRSe-TIME, struggle to achieve strong predictive performance. These results emphasize that while SPaRSe-TIME excels in structured time series with clear temporal components, its advantages diminish in highly stochastic settings where saliency dominates, and trend information is weak.

\subsection{Results on UCI Air Quality Dataset \cite{air_quality_360}}

\subsubsection{Quantitative Results}

Table~\ref{tab:air_quality_results} presents the performance comparison across all models on the UCI Air Quality dataset.

\begin{table}[t]
\centering
\caption{Performance comparison on UCI Air Quality dataset}
\label{tab:air_quality_results}
\begin{tabular}{lccc}
\hline
Model & MAE $\downarrow$ & RMSE $\downarrow$ & $R^2$ $\uparrow$ \\
\hline
RNN          & 0.313 & 0.526 & -0.970 \\
GRU          & 0.300 & 0.582 & -1.409 \\
LSTM         & 0.355 & 0.690 & -2.386 \\
Transformer  & \textbf{0.088} & \textbf{0.364} & \textbf{0.057} \\
SPaRSe-TIME  & 0.123 & 0.401 & -0.142 \\
\hline
\end{tabular}
\end{table}

The Transformer achieves the best overall performance on the UCI Air Quality dataset, significantly outperforming other models across all metrics. It is the only model that attains a positive $R^2$ score, indicating its ability to capture complex temporal dependencies in the data. SPaRSe-TIME achieves competitive performance, outperforming all recurrent baselines (RNN, GRU, LSTM) in terms of MAE and RMSE, but falls short of the Transformer. The results suggest that the air quality dataset exhibits complex, nonlinear temporal patterns that are better captured by attention-based models. Unlike datasets with clear decomposition structures, such as weather or household energy data, the air quality data contains intricate dependencies across multiple variables and time scales. While SPaRSe-TIME benefits from its structured decomposition, its performance is limited in this setting due to the weaker separability of saliency, memory, and trend components. The relatively low $R^2$ score indicates that the decomposition may not fully capture the underlying dynamics of air pollutant concentrations. Nevertheless, SPaRSe-TIME demonstrates improved performance over traditional recurrent models, highlighting its ability to model temporal structure more effectively than standard sequence models.

\subsection{Computational Efficiency Analysis}

\subsubsection{Model Complexity}

Table~\ref{tab:computational_results} compares the computational efficiency of all models in terms of parameter count, floating-point operations (FLOPs), and inference time on the household dataset.

\begin{table}[t]
\centering
\caption{Computational efficiency comparison across models}
\label{tab:computational_results}
\begin{tabular}{lccc}
\hline
Model & Params (K) & FLOPs (G) & Time (ms) \\
\hline
RNN          & 4.6   & 0.03 & 0.006 \\
GRU          & 13.69 & 0.09 & 0.008 \\
LSTM         & 18.24 & 0.12 & 0.008 \\
Transformer  & 43.35 & 0.25 & 0.030 \\
\textbf{SPaRSe-TIME} & \textbf{1.09} & \textbf{0.01} & \textbf{0.005} \\
\hline
\end{tabular}
\end{table}

The proposed SPaRSe-TIME achieves the lowest parameter count and computational cost, requiring only 1.09K parameters and 0.01G FLOPs. Despite its lightweight design, it also attains the fastest inference time, highlighting its efficiency. In contrast, Transformer-based models incur significantly higher computational overhead, while recurrent models exhibit moderate complexity but remain less efficient than the proposed approach.

\subsection{Ablation Study}

An ablation experiment is performed to systematically evaluate the contribution of each component of the proposed SParSe-TIME across two representative datasets: the Household Power Consumption dataset and the Weather dataset.

\begin{table}[t]
\centering
\caption{Ablation study across datasets}
\label{tab:ablation_combined}
\begin{tabular}{lcccccc}
\hline
 & \multicolumn{3}{c}{Household} & \multicolumn{3}{c}{Weather} \\
\cline{2-4} \cline{5-7}
Configuration & MAE $\downarrow$ & RMSE $\downarrow$ & $R^2$ $\uparrow$ & MAE $\downarrow$ & RMSE $\downarrow$ & $R^2$ $\uparrow$ \\
\hline
Full (S+M+G)    & \textbf{0.398} & \textbf{0.554} & \textbf{0.503} & \textbf{0.457} & \textbf{0.616} & \textbf{0.595} \\
No Saliency     & 0.420 & 0.578 & 0.459 & 0.457 & 0.616 & 0.594 \\
No Memory       & 0.799 & 0.922 & -0.377 & 0.577 & 0.739 & 0.416 \\
No Trend        & 0.428 & 0.575 & 0.465 & 0.661 & 0.779 & 0.350 \\
Only Memory     & 0.452 & 0.600 & 0.418 & 0.648 & 0.764 & 0.376 \\
Only Saliency   & 0.893 & 1.023 & -0.694 & 0.833 & 0.995 & -0.058 \\
Only Trend      & 0.839 & 0.960 & -0.491 & 0.566 & 0.727 & 0.436 \\
\hline
\end{tabular}
\end{table}

The results demonstrate that combining all components (S+M+G) consistently yields the best performance across both datasets, confirming the effectiveness of the proposed decomposition framework. A clear pattern emerges regarding component importance:

\paragraph{Memory Dominance}
Memory is the most critical component across both datasets. Removing memory results in the largest performance degradation, particularly on the household dataset, where local temporal dependencies are strong. This indicates that short-term dynamics play a fundamental role in time series prediction.

\paragraph{Trend Dependency}
The importance of the trend component is dataset-dependent. On the weather dataset, removing trend leads to a significant drop in performance, highlighting the presence of strong long-term temporal patterns. In contrast, its impact on the household dataset is comparatively smaller.

\paragraph{Saliency Contribution}
The saliency component contributes moderately and is less critical than memory and trend. Removing saliency has minimal impact on the weather dataset and only a moderate effect on the household dataset, suggesting that high-frequency variations are less informative for prediction.

\paragraph{Single Component Limitations}
Using any single component alone leads to substantially worse performance, confirming that no individual temporal aspect is sufficient to model complex time series dynamics. In particular, saliency-only and trend-only configurations perform poorly, emphasizing the need for multi-scale modeling.

These findings highlight that SPaRSe-TIME effectively integrates complementary temporal components. While memory provides a strong baseline representation, incorporating saliency and trend enables the model to adapt to dataset-specific characteristics, resulting in improved performance and robustness across diverse time series domains.

\subsection{Interpretability Analysis}

To understand how SPaRSe-TIME adapts to different time series characteristics, learned component weights ($\alpha$) are analyzed corresponding to saliency, memory, and trend across all datasets.

\begin{figure}[h]
\centering
\includegraphics[width=\linewidth]{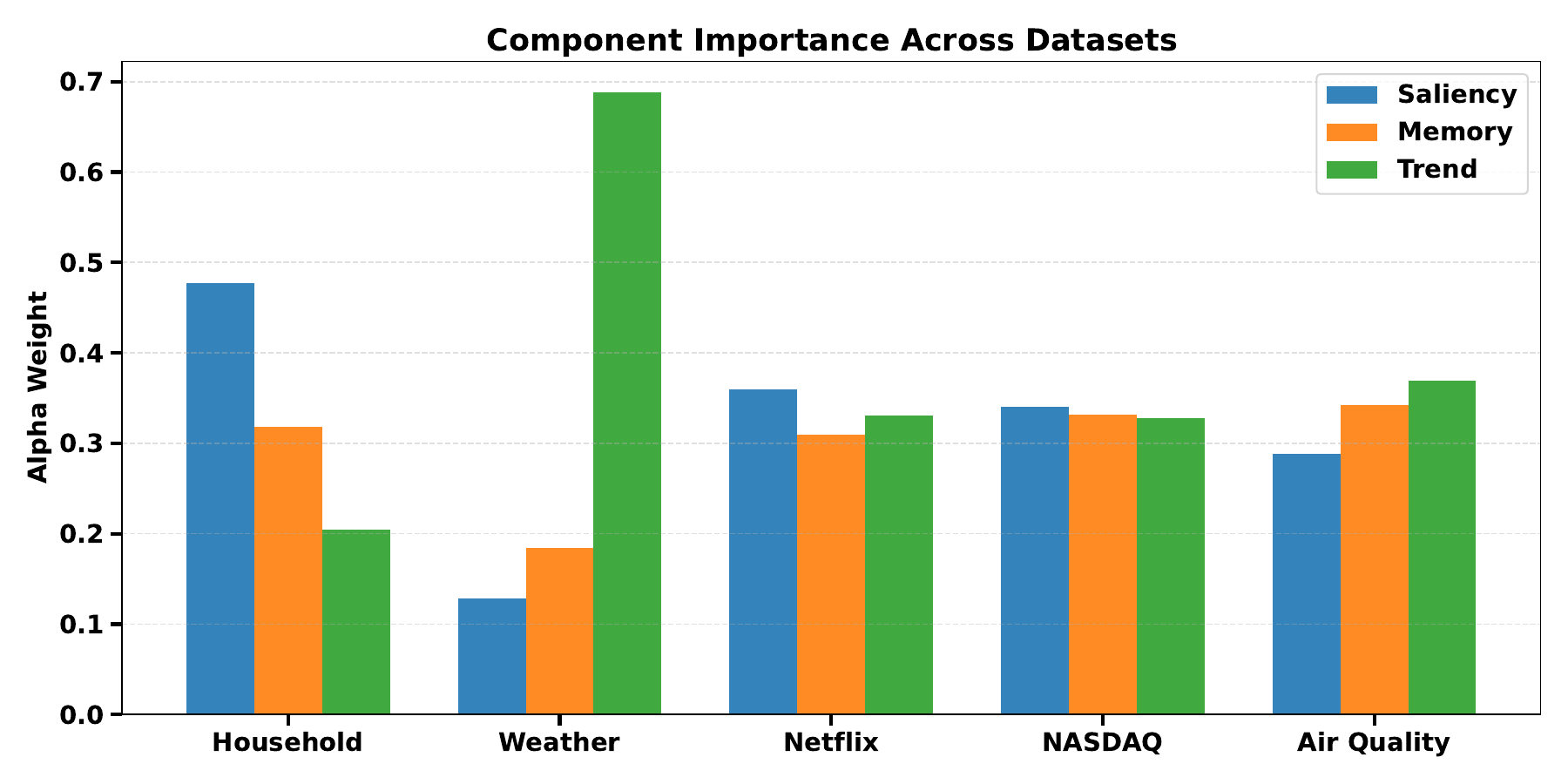}
\caption{Learned component weights ($\alpha$) across datasets.}
\label{fig:alpha_comparison}
\end{figure}

Figure~\ref{fig:alpha_comparison} illustrates the relative importance of each component for different datasets. A clear variation in component dominance can be observed, highlighting the adaptive nature of the proposed model.

\paragraph{Household Dataset}
The model assigns the highest weight to the saliency component, indicating that short-term fluctuations and rapid changes play a dominant role in power consumption patterns. Memory contributes moderately, while trend has a relatively smaller influence.

\paragraph{Weather Dataset}
In contrast, the trend component dominates significantly, receiving the largest weight among all datasets. This reflects the strong presence of long-term temporal patterns and seasonal effects in weather data. Saliency contributes minimally, indicating that high-frequency variations are less informative.

\paragraph{Netflix Dataset}
The weights are relatively balanced across all three components, suggesting that the dataset contains a mix of short-term fluctuations, local dependencies, and weak trends. This balanced structure aligns with the noisy and partially structured nature of stock price data.

\paragraph{NASDAQ Dataset}
Similar to the Netflix dataset, the weights are nearly uniform, indicating no single dominant temporal component. This suggests that the dataset is highly stochastic, with limited exploitable temporal structure, which explains the overall difficulty in achieving strong predictive performance.

\paragraph{Air Quality Dataset}
The model assigns higher importance to memory and trend components, indicating that both local dependencies and gradual temporal changes are important in modeling air pollutant dynamics. Saliency plays a comparatively smaller role.

These results demonstrate that SPaRSe-TIME does not rely on a fixed decomposition strategy but instead adapts dynamically to the underlying characteristics of each dataset. The variation in $\alpha$ values provides strong evidence that the model effectively captures dataset-specific temporal structures, enhancing both interpretability and performance. This adaptive behavior is a key advantage over traditional models, which learn implicit representations without providing insight into the relative importance of different temporal components.

\section{Discussion}

The experimental results demonstrate that \textbf{SPaRSe-TIME} is highly effective in modeling structured time series data, particularly in domains where temporal dynamics can be decomposed into meaningful components such as saliency, memory, and trend. Across the household power consumption and weather datasets, the proposed model consistently outperforms all baseline methods, achieving lower error metrics and higher $R^2$ scores. These improvements can be attributed to the model’s ability to explicitly disentangle temporal structures and focus on informative components, rather than uniformly processing all time steps. The interpretability analysis further confirms that the model adapts its component weights according to dataset characteristics, emphasizing saliency in volatile signals and trend in smooth, seasonal data.

However, the results also reveal important limitations. On highly noisy and stochastic datasets such as Netflix and NASDAQ, all models—including SPaRSe-TIME—achieve low or negative $R^2$ scores, indicating limited predictability. In these settings, the assumption of structured temporal decomposition becomes less valid, as the signal lacks clear low-frequency trends or stable memory patterns. As a result, the advantages of the proposed decomposition diminish, and the model behaves similarly to baseline approaches. In particular, SPaRSe-TIME underperforms compared to the Transformer on the NASDAQ dataset, suggesting that global attention mechanisms may better capture weak and distributed dependencies in highly stochastic environments. Furthermore, on the UCI Air Quality dataset, Transformer-based models outperform SPaRSe-TIME, highlighting another limitation of the proposed framework. Air quality data exhibits complex, nonlinear interactions across variables and time scales, which are not fully captured by the current decomposition into saliency, memory, and trend components. While SPaRSe-TIME improves over recurrent models, its relatively simple linear projections and component-wise aggregation may limit its capacity to model intricate cross-dimensional dependencies. :contentReference[oaicite:0]{index=0}

From a modeling perspective, several limitations can be identified. First, the saliency operator relies on first-order temporal differences, which may not adequately capture higher-order or context-dependent importance. Second, the low-rank memory representation assumes that temporal dynamics lie in a fixed low-dimensional subspace, which may not hold for highly nonlinear or regime-switching processes \cite{KishoreKumar02112017}. Third, the trend component is modeled using simple smoothing operators, which may be insufficient for capturing complex long-term dependencies \cite{4469947}. Finally, the interaction between components is relatively shallow, relying on linear transformations and weighted aggregation.

These observations suggest several promising directions for future research. One potential extension is to incorporate \emph{adaptive or learned saliency mechanisms}, allowing the model to identify informative time steps based on richer contextual features rather than simple gradients \cite{balestra2023consistencyrobustnesssaliencyexplanations}. Another direction is to replace the fixed low-rank projection with \emph{dynamic or nonlinear subspace learning}, enabling the memory component to better capture evolving temporal structures \cite{6514578}. Additionally, integrating \emph{lightweight attention or graph-based interactions} could enhance the model’s ability to capture cross-variable dependencies while preserving computational efficiency. Finally, more expressive trend modeling techniques, such as multi-scale decomposition or frequency-domain filtering, could improve performance on datasets with complex long-term dynamics \cite{ZHAO2026100883,10.1145/3783986}.

Overall, the results indicate that SPaRSe-TIME provides a compelling trade-off between efficiency, interpretability, and predictive performance. It is particularly well-suited for structured time series with clear temporal components, while its limitations highlight the challenges of modeling highly stochastic and strongly nonlinear systems. These findings underscore the importance of combining structured decomposition with flexible modeling mechanisms in future time series architectures.

\section{Conclusion}

This work introduced \textbf{SPaRSe-TIME}, a novel and computationally efficient framework for time series prediction based on a structured decomposition of temporal signals into saliency, memory, and trend components. By reformulating temporal modeling as a projection onto informative subspaces, the proposed approach enables selective processing of time series data, in contrast to conventional models that treat all time steps uniformly. Extensive experiments across diverse real-world datasets demonstrate that SPaRSe-TIME achieves competitive or superior performance compared to established baselines, while significantly reducing computational complexity. The model is particularly effective in structured time series domains, where distinct temporal components such as trends and local dependencies are present. Moreover, the framework provides inherent interpretability through learnable component weights, offering insights into how different temporal structures contribute to predictions. However, the results also highlight limitations in highly stochastic and complex multivariate settings, where the assumption of separable temporal components becomes less effective. These findings suggest that future work should focus on enhancing the flexibility of component modeling and incorporating richer interaction mechanisms. Overall, SPaRSe-TIME presents a principled and efficient alternative to traditional sequence modeling approaches, bridging the gap between interpretability and performance, and providing a promising direction for scalable and explainable time series learning.

\bibliographystyle{unsrtnat}
\bibliography{references}  






\end{document}